\documentclass[twocolumn,showpacs,preprintnumbers,amsmath,amssymb,prl,floatfix]{revtex4}

\usepackage{graphicx}
\usepackage{dcolumn}
\usepackage{bm}
\usepackage{color}

\begin{document}

\title{
Quantum Phase Diagram of the Triangular-Lattice {\it XXZ} Model in a Magnetic Field
}

\author{Daisuke Yamamoto$^{1}$}
\author{Giacomo Marmorini$^{1,2}$}
\author{Ippei Danshita$^{3}$}
\affiliation{
{$^1$Condensed Matter Theory Laboratory, RIKEN, Saitama 351-0198, Japan}
\\
{$^2$Research and Education Center for Natural Sciences, Keio University, Kanagawa 223-8521, Japan}
\\
{$^3$Yukawa Institute for Theoretical Physics, Kyoto University, Kyoto 606-8502, Japan\\
and Computational Condensed Matter Physics Laboratory, RIKEN, Saitama 351-0198, Japan}
}
\date{\today}
\begin{abstract}
The triangular lattice of $S=1/2$ spins with {\it XXZ} anisotropy is a ubiquitous model for various frustrated systems in different contexts. We determine the quantum phase diagram of the model in the plane of the anisotropy parameter and the magnetic field by means of a large-size cluster mean-field method with a scaling scheme. 
We find that quantum fluctuations break up the nontrivial continuous degeneracy into two first-order phase transitions. In between the two transition boundaries, the degeneracy lifting results in the emergence of a new coplanar phase not predicted in the classical counterpart of the model. We suggest that the quantum phase transition to the nonclassical coplanar state can be observed in triangular-lattice antiferromagnets with large easy-plane anisotropy or in the corresponding optical-lattice systems.  
\end{abstract}
\pacs{75.10.Jm,75.45.+j,75.30.Kz}
\maketitle

\emph{Introduction.}--- Geometric frustration arises when local interaction energies cannot be simultaneously minimized due to lattice geometry, resulting in a large ground-state degeneracy~\cite{toulouse-77,moessner-06}. A variety of unconventional phenomena generated by frustration have been a fascinating and challenging subject in modern condensed matter physics. In particular, frustrated spin systems are a promising place to explore exotic states of matter such as noncolinear antiferromagnetic order~\cite{capriotti-99,zheng-06,white-07,harada-12}, order-by-disorder selection to form magnetization plateaus~\cite{chubokov-91,farnell-09,sakai-11,nishimoto-13}, spin liquid~\cite{balents-10,han-12}, and lattice supersolidity~\cite{wang-09,jiang-09,heidarian-10}. However, established theories and numerical simulations often encounter serious difficulties including the notorious minus-sign problem~\cite{suzuki-93} in dealing with frustrated systems. Thus a reliable investigation for frustrated magnetism has been limited mainly to classical spins~\cite{kawamura-85,miyashita-86,seabra-11}, the $SU(2)$-symmetric point of the model~\cite{capriotti-99,zheng-06,white-07,harada-12}, or (quasi-)one-dimensional systems~\cite{chen-13}.

In this Letter, we demonstrate a possible way to overcome the problem by determining the ground-state phase diagram of frustrated quantum spins on the 2D triangular lattice over a wide range of magnetic field and exchange anisotropy. This system has also been attracting great physical interest from the experimental side since the latest developments in magnetic materials and ultracold gases have resolved technical difficulties to realize ideal 2D frustrated systems. Specifically, the compound Ba$_3$CoSb$_2$O$_9$ has been reported very recently~\cite{shirata-12,zhou-12,susuki-13,koutroulakis-13} as the first example of ideal triangular-lattice antiferromagnet with spatially isotropic couplings and no Dzyaloshinsky-Moriya interactions. In this compound, the effective $S=1/2$ spins of Co$^{2+}$ ions form a regular triangular lattice unlike other known (distorted) materials such as Cs$_2$CuCl$_4$~\cite{coldea-01}, Cs$_2$CuBr$_4$~\cite{ono-03,fortune-09}, and $\kappa$-(BEDT-TTF)$_2$Cu$_2$(CN)$_3$~\cite{shimizu-03}. The magnetization process of the single-crystal samples has shown a strong dependence on the magnetic field direction~\cite{susuki-13}, which indicates the existence of the anisotropy between the in-plane ({\it XY}) and out-of-plane (Ising) exchange interactions in spin space, known as {\it XXZ} anisotropy. To properly explain the observed magnetization anomalies, it is necessary to take into account the exchange anisotropy and quantum fluctuations for arbitrary field. Furthermore, considerable advances have also been made in the direction of simulating magnetism with ultracold atomic or molecular gases in a periodic optical potential~\cite{trotzky-08,simon-11,struck-11,greif-13}. A frustrated {\it XY} system has indeed been realized recently~\cite{struck-11} by dynamically inverting the sign of the hopping integral~\cite{eckardt-05,lignier-07} of bosonic atoms in a triangular optical lattice~\cite{becker-10}. In optical lattices, an Ising-type coupling can be introduced by finite-range repulsion, e.g., dipole-dipole interactions~\cite{yamamoto-12,pollet-10,yamamoto-13}, while the {\it XY} coupling comes from the hopping. Thus the {\it XXZ} anisotropy is widely controllable in such a system.

\begin{figure}[t]
\includegraphics[scale=0.19]{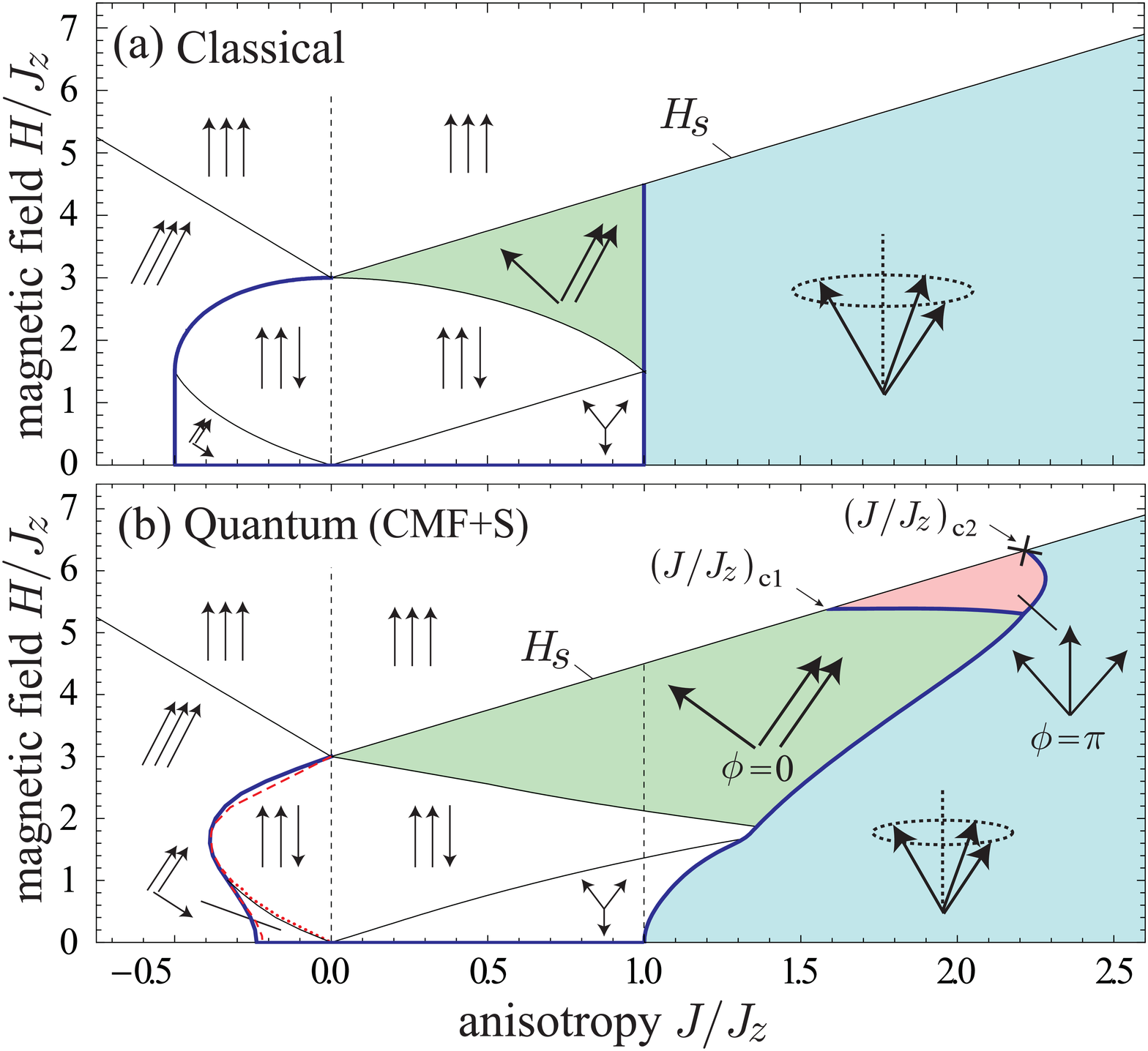}
\caption{\label{fig1}
Ground-state phase diagram of the spin-1/2 triangular-lattice {\it XXZ} model from (a) the classical and (b) CMF+S analyses ($J_z>0$). 
The thick blue (thin black) solid curves correspond to first- (second-) order transitions. 
{The latest QMC data~\cite{wessel-05,bonnes-11} are shown by the red dashed (first-order) and dotted (second-order) curves. The symbol ($\times$) is the value from the dilute Bose-gas expansion.} 
}
\end{figure}

In connection with the ongoing experiments, we report a theoretical prediction of the quantum phase diagram of the spin-1/2 frustrated {\it XXZ} system on the triangular lattice with the following Hamiltonian~\cite{nishimori-86}: 
\begin{eqnarray}
\hat{\mathcal{H}}&\!\!\!=\!\!\!&
J\!\sum_{\langle i,j\rangle}\!\Big(\hat{S}_i^x\hat{S}_j^x+\hat{S}_i^y\hat{S}_j^y\Big)\!+\!J_z\!\sum_{\langle i,j\rangle}\hat{S}_i^z\hat{S}_j^z\!-\!H\!\sum_{i}\hat{S}^z_{i},
\label{hamiltonian}
\end{eqnarray}
where the sum $\sum_{\langle i,j\rangle}$ runs over nearest-neighbor sites. The spin-1/2 {\it XXZ} model is also an effective model describing spin-dimer compounds such as Ba$_3$Mn$_2$O$_8$, in which the isotropic couplings can induce large effective {\it XXZ} anisotropy~\cite{samulon-08}, and binary mixtures of atomic gases in an optical lattice~\cite{kuklov-03, altman-03}. Despite the broad relevance and the apparent simplicity of the model (\ref{hamiltonian}), its quantum phase diagram in the frustrated regime ($J, J_z > 0$) remains unrevealed mainly because the quantum Monte Carlo (QMC) method suffers from the minus-sign problem. Here, we avoid the usual difficulties for frustrated systems by employing the large-size cluster mean-field method combined with a scaling scheme (CMF+S) established recently in Ref.~\cite{yamamoto-12-2} and determine the complete quantum phase diagram in the whole plane of the anisotropy $-\infty < J/J_z<\infty$ and the magnetic field $H/J_z$ for $J_z>0$ with a high degree of accuracy [see Fig.~\ref{fig1}]. We show that quantum fluctuations drastically change the phase diagram from the classical one. In particular, we find that the nontrivial continuous degeneracy at $J/J_z=1$ breaks up into two first-order transitions at strong fields due to the quantum effects, and a non-classical coplanar state emerges between the two transitions. 
We complement the analysis with the dilute Bose gas expansion~\cite{nikuni-95} near the saturation field and express the first-order transitions in terms of the magnon Bose-Einstein condensation (BEC).
We also discuss a translation of the results into the bosonic language with optical-lattice experiments in mind. 

\begin{figure}[t]
\includegraphics[scale=0.36]{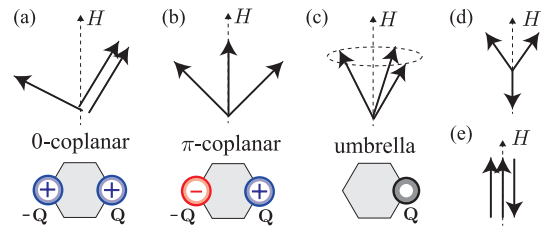}
\caption{\label{fig2}
Five types of spin configurations for $J/J_z>0$. The sets of three arrows represent each spin angle on the three sublattices. The lower illustrations in (a)-(c) depict the corresponding distributions of magnon BECs at the corners of the hexagonal first Brillouin zone. 
}
\end{figure}

\emph{Classical phase diagram.}--- In Fig.~\ref{fig1}(a), we show the phase diagram obtained by the classical-spin ($S=\infty$) analysis~\cite{miyashita-86,murthy-97} as reference to be compared with the quantum case. 
For positive easy-axis anisotropy $0<J/J_z<1$, one finds three different states with the three-sublattice $\sqrt{3}\times\sqrt{3}$ structure below the saturation field $H_{s}=3J/2+3J_z$: low- and high-field coplanar states depicted in Figs.~\ref{fig2}(d) and \ref{fig2}(a) and a collinear up-up-down state in Fig.~\ref{fig2}(e). For easy-plane anisotropy $J/J_z>1$, the so-called umbrella state in Fig.~\ref{fig2}(c) appears. We will discuss quantum effects on the classical ground state by means of the dilute Bose-gas~\cite{nikuni-95} and CMF+S~\cite{yamamoto-12-2} approaches. It is of particular interest how the ground-state degeneracy along the line of $J/J_z=1$~\cite{kawamura-85} is lifted. 

%

\emph{Dilute Bose-gas expansion.}---The quantum magnetic structures just below $H_{s}$ can be semianalytically studied using the dilute Bose-gas expansion~\cite{nikuni-95,batuev-84}, in which first the spin model (\ref{hamiltonian}) is rewritten in the hard-core boson (magnon) representation: $\hat{S}_i^z=1/2-\hat{a}_i^\dagger\hat{a}_i$ and $\hat{S}_i^+=\hat{a}_i$. For the triangular lattice, the magnons $\hat{a}_{\bf k}$ in the Fourier space can condense at either or both of the two independent minima of the single-particle energy, which are located at the corners ${\bf k}=\pm {\bf Q}\equiv\pm(4\pi/3,0)$ of the hexagonal first Brillouin zone.  
For $0< H_{s}-H \ll H_{s}$, the ground-state energy per site up to fourth order in the magnon BEC order parameters $\psi_{\pm {\bf Q}}\equiv \langle \hat{a}_{\pm{\bf Q}} \rangle$ is given by
\begin{eqnarray}
E_0/M&\!\!=\!\!&-(H_{s}-H)\left(|\psi_{{\bf Q}}|^2+|\psi_{-{\bf Q}}|^2\right)\nonumber\\&&\!\!+\mathit{\Gamma}_1\left(|\psi_{{\bf Q}}|^4+|\psi_{-{\bf Q}}|^4\right)/2\!+\!\mathit{\Gamma}_2|\psi_{{\bf Q}}|^2|\psi_{-{\bf Q}}|^2.
\label{ene0}
\end{eqnarray}
The degeneracy in the relative phase $\phi=\arg (\psi_{{\bf Q}}/\psi_{-{\bf Q}})$ between the two BECs can be lifted by the higher-order term $2\mathit{\Gamma}_3|\psi_{{\bf Q}}|^3|\psi_{-{\bf Q}}|^3\cos 3\phi$. More details of Eq.~(\ref{ene0}) and the effective interactions $\mathit{\Gamma}_1$, $\mathit{\Gamma}_2$, and $\mathit{\Gamma}_3$ are presented in the Supplemental Material~\cite{EPAPS}. The ordering vectors $\pm {\bf Q}$ identify a three-sublattice structure consistent with the classical-spin analysis. Minimizing the ground-state energy, we obtain the following three types of solution:

(i) $\mathit{\Gamma}_1>\mathit{\Gamma}_2$ and $\mathit{\Gamma}_3<0$: $|\psi_{{\bf Q}}|=|\psi_{-{\bf Q}}|\neq 0$, $\phi=0$;

(ii) $\mathit{\Gamma}_1>\mathit{\Gamma}_2$ and $\mathit{\Gamma}_3>0$: $|\psi_{{\bf Q}}|=|\psi_{-{\bf Q}}|\neq 0$, $\phi=\pi$;

(iii) $\mathit{\Gamma}_1<\mathit{\Gamma}_2$: $|\psi_{{\bf Q}}|\neq 0$ and $|\psi_{-{\bf Q}}|= 0$ (or vice versa). \\
Since the double-BEC solutions with (i) $\phi=0$  and (ii) $\phi=\pi$ correspond to the two different coplanar states in Figs.~\ref{fig2}(a) and \ref{fig2}(b)~\cite{nikuni-95}, we refer to them as the ``0-coplanar'' and ``$\pi$-coplanar'' states. The single-BEC solution (iii) is translated into the umbrella state in Fig.~\ref{fig2}(c).

\begin{figure}[b]
\includegraphics[scale=0.39]{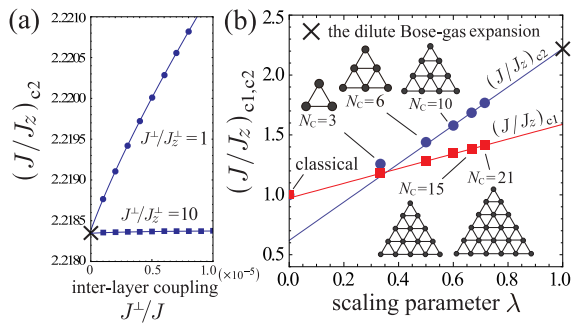}
\caption{\label{fig3} (a) Coplanar-umbrella phase boundary $(J/J_z)_{c2}$ just below the saturation field as a function of the interlayer coupling strength obtained from $\mathit{\Gamma}_1=\mathit{\Gamma}_2$. We display the cases of the isotropic (circles) and {\it XY}-type (squares) antiferromagnetic interlayer couplings as examples. (b) Cluster-size scaling of the CMF data for the phase boundaries $(J/J_z)_{{c}1}$ between 0- and $\pi$-coplanar phases as well as $(J/J_z)_{{c}2}$ just below the saturation field. }
\end{figure}

We calculate the coplanar-umbrella phase boundary $(J/J_z)_{c2}$ from the condition $\mathit{\Gamma}_1=\mathit{\Gamma}_2$. In 2D systems, $\mathit{\Gamma}_1$ and $\mathit{\Gamma}_2$ vanish due to the infrared singularity in loop integrals~\cite{fisher-88}. Therefore, we introduce interlayer {\it XXZ} couplings $J^\perp,J_z^\perp$ {as regulators}, and then take the limit of $J^\perp,J_z^\perp\rightarrow 0$~\cite{EPAPS}. The value of $(J/J_z)_{c2}$ converges to $2.218$ regardless of the sign and ratio of $J^\perp$ and $J_z^\perp$ {(or, in other words, independently of the details of the regularization)} [see Fig.~\ref{fig3}(a)]. This means that the region of coplanar states is extended toward the rather large easy-plane anisotropy side due to the quantum effects [see the symbol ($\times$) in Fig.~\ref{fig1}(b)]. Even for $0< H_{s}-H \ll H_{s}$, the dilute Bose-gas expansion has not been able to determine which coplanar state ($\phi=0$ or $\pi$) emerges, because the calculation of $\mathit{\Gamma}_3$ is practically difficult~\cite{nikuni-95}. We will see below that the CMF+S analysis unambiguously answers this long-standing question, first raised in Ref.~\cite{nikuni-95}.

\emph{Entire quantum phase diagram.}---The complete quantum phase diagram for an arbitrary field is numerically determined by the use of the CMF+S method. We perform the exact diagonalization of a cluster system of $N_{C}$ spins after the standard mean-field decoupling of the interactions between the edge and outside spins~\cite{yamamoto-12-2}. Although we treat only static mean fields unlike the (cluster) dynamical mean-field approximation~\cite{kotliar-01,anders-10}, we can deal with a large-size cluster which gives the possibility to take the infinite cluster-size limit~\cite{yamamoto-12-2,yamamoto-13,luhmann-13}. Here, we use the series of the clusters that consist of up to $N_{C}=21$ spins and self-consistently calculate $m^{\alpha}_{\mu}\equiv \langle \hat{S}^{\alpha}_{i_{\mu}}\rangle$ ($\alpha=x,y,z$) considering all possible spin structures under the three-sublattice ansatz ($\mu={A},{B},{C}$). 
We find that the data for the phase boundaries obtained by the three largest clusters produce a linear extrapolation line with the scaling parameter $\lambda\equiv N_{B}/(N_{C} z/2)$ [see Fig.~\ref{fig3}(b)], which allows us to determine the phase diagram of the frustrated spin model~(\ref{hamiltonian}) in a quantitatively reliable way. Here, $N_{B}$ is the number of bonds within the cluster and $z=6$ is the coordination number of the triangular lattice.

The quantum phase diagram is shown in Fig.~\ref{fig1}(b). We see that the positive (frustrated) $J/J_z$ side is drastically changed from the classical one. The collinear up-up-down state is extended by quantum effects, which causes a plateau at one-third of the saturation magnetization in the magnetization process even for $J/J_z\geq 1$. The coplanar states are also significantly extended toward the easy-plane side for strong fields. Just below the saturation field, the scaled value of the coplanar-umbrella boundary is $(J/J_z)_{{ c}2}=2.220$ [see Fig.~\ref{fig3}(b)], which is in good agreement with the value $2.218$ from the dilute Bose-gas expansion. Of particular interest is the emergence of a new phase not predicted in the classical counterpart of the model for large easy-plane anisotropy $1.6\lesssim J/J_z\lesssim 2.3$ and strong fields $H/H_{s}\gtrsim 0.84$ [red region in Fig.~\ref{fig1}(b)] as a result of a novel quantum lifting mechanism (explained below). The spin structure of the nonclassical state is given by $m^{z}_{A}\neq m^{z}_{ B}= m^{z}_{ C}$ and $m^{x}_{ A}=0,~m^{x}_{ B}=- m^{x}_{ C}$ when the ordering plane is the $xz$ plane ($m^{y}_\mu=0$). This is indeed the $\pi$-coplanar state shown in Fig.~\ref{fig2}(b). On the other hand, $m^{z}_{ A}=m^{z}_{B}\neq m^{z}_{ C}$ and $m^{x}_{ A}=m^{x}_{ B}\neq m^{x}_{ C}$ in the 0-coplanar state (green region). The 0-$\pi$ transition point just below the saturation field is extrapolated to $(J/J_z)_{{ c}1}=1.588$, at which the sign of $\mathit{\Gamma}_3$ should change. The total transverse magnetization is nonvanishing in the 0-coplanar state ($2m^{x}_{ A}+m^{x}_{ C}\neq 0$)~\cite{nishimori-86}, whereas it is zero in the $\pi$-coplanar state.

The quantum phase diagram does not include any disordered phase, i.e., spin liquid. The two end points of the plateau at $J/J_z=1$ are given by $H_{{ c}1}/J_z= 1.345$ and $H_{{ c}2}/J_z=2.113$, which are consistent with the coupled cluster method~\cite{farnell-09} and the exact diagonalization with periodic boundary conditions~\cite{sakai-11}. {Moreover, our result gives good agreement with the QMC data~\cite{wessel-05,bonnes-11} (red curves) in the negative $J/J_z$ side including the order of the transitions~\cite{footnote,bonnes-11,zhang-11}. In particular, the phase transition point at $H=0$, $(J/J_z)_0=-0.238$, agrees with the known numerical data, $(J/J_z)_0\approx -0.23$ - $-0.21$~\cite{wessel-05,bonnes-11,zhang-11,OpticalLattice} (see the comparison table in Ref.~\cite{EPAPS})}, which indicates high accuracy of the CMF+S analysis on the current problem.

\begin{figure}[t]
\includegraphics[scale=0.26]{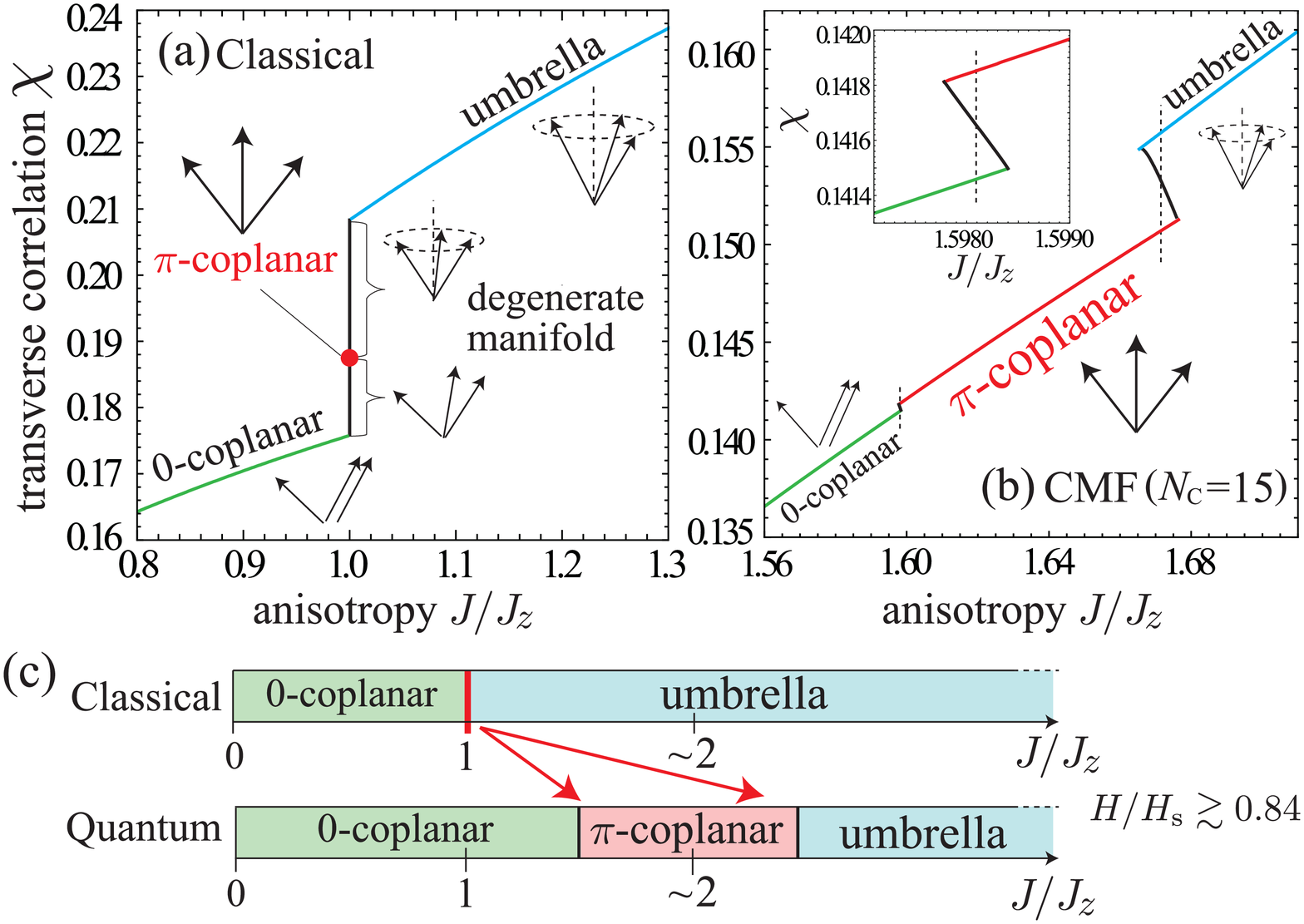}
\caption{\label{fig4} (a) Classical solution of $\chi$ as a function of $J/J_z$ for $H/J_z=3$. {(b) Quantum degeneracy lifting obtained by the CMF analysis ($H/J_z=4.5$). The vertical dashed lines mark the first-order transition points determined by the Maxwell construction. The inset is the enlarged view around the weak first-order 0-$\pi$ transition.} The phase diagrams in (c) show the quantum breakup of the continuous degeneracy into two first-order transitions for $H/H_{ s}\gtrsim 0.84$. }
\end{figure}

\emph{Degeneracy-lifting mechanism.}---In Fig.~\ref{fig4}(a), we plot the classical solution curve in the plane of the conjugate thermodynamic variables: $J/J_z$ and the transverse nearest-neighbor correlation $\chi\equiv  -\sum_{\langle i,j\rangle}\langle \hat{S}_i^x\hat{S}_j^x+\hat{S}_i^y\hat{S}_j^y\rangle/M$. At $J/J_z=1$, there is a nontrivial continuous degeneracy of ground states in which the classical-spin vectors satisfy $\bm{S}_{ A}+\bm{S}_{ B}+\bm{S}_{ C}=(0,0,H/3J)$ with $|\bm{S}_\mu|=1/2$~\cite{kawamura-85}. Figure~\ref{fig4}(b) illustrates the peculiar mechanism of the quantum degeneracy lifting. The quantum fluctuations select the $\pi$-coplanar state out of the continuous manifold of the classical ground states. In the solution curve, the point of the $\pi$-coplanar state shown in Fig.~\ref{fig4}(a) is extended to a finite section in Fig.~\ref{fig4}(b). All of the other intermediate states form two separate sections of the solution curve with negative slope (negative ``susceptibility''), which indicates the instability of those states. As a result, the classical ground-state degeneracy is broken up into two first-order transitions [see Fig.~\ref{fig4}(c)].

This novel degeneracy-lifting mechanism is sharply different from the known cases. For example, the square-lattice {\it XXZ} model also possesses a classical continuous degeneracy at the boundary of the spin-flop transition from the N\'eel to canted antiferromagnetic phase~\cite{holtschneider-07,kohno-97}. However, all of the intermediate states in the degenerate manifold are destabilized and only a single first-order transition is induced by the quantum effects~\cite{holtschneider-07,kohno-97,batrouni-00} (see Ref.~\cite{EPAPS} for the direct comparison with Fig.~\ref{fig4}). The same behavior also appears in certain bosonic systems such as spin-2 BECs at the transition boundaries to nematic phases~\cite{phuc-13}. In contrast, in the present model a specific intermediate state is chosen by quantum fluctuations from the degenerate manifold and occupies a finite region of the quantum phase diagram, whereas it does not appear in the classical one. 

%

\emph{Remarks on experiments.}---
In the experiment of Ref.~\cite{susuki-13} on Ba$_3$CoSb$_2$O$_9$, the magnetization curve exhibits a cusp at $H\approx H_{ s}/3$ for magnetic fields parallel to the $c$ axis and a clear plateau is not detected. This can be understood within the phase diagram in Fig.~\ref{fig1}(b) if the anisotropy is as large as $J/J_z\approx 1.3$. 
The authors in Ref.~\cite{susuki-13} have conjectured that a magnetization anomaly in Ba$_3$CoSb$_2$O$_9$ under transverse magnetic field $H\perp c$ may correspond to the 0-$\pi$ transition of coplanar states, which is still controversial~\cite{koutroulakis-13}. Moreover, the first-order 0-$\pi$ transition for $H\parallel c$ is expected to be observed as a jump in the magnetization process by synthesizing a family material with larger easy-plane anisotropy $1.6\lesssim J/J_z\lesssim 2.3$ or by tuning $J/J_z$ with pressure~\cite{ruegg-08} in spin-dimer compounds such as Ba$_3$Mn$_2$O$_8$~\cite{samulon-08}.

In the context of cold atomic or molecular systems, one could prepare the spin-1/2 {\it XXZ} system using, e.g., dipolar bosons with strong on-site repulsions in a triangular optical lattice~\cite{yamamoto-12,pollet-10,yamamoto-13}. The frustrated regime $J,J_z>0$ could be accessed by the latest techniques such as a fast oscillation of the lattice~\cite{struck-11,eckardt-05,lignier-07}. In the language of the hard-core boson, $1/2-m^{z}_\mu$ and $[(m^{x}_\mu)^2+(m^{y}_\mu)^2]^{1/2}$ correspond to the sublattice density filling and the sublattice BEC order parameter, respectively~\cite{matsuda-70}. Therefore, the 0-coplanar state is regarded as a lattice supersolid (SS) state. Although the bosonic counterpart of the $\pi$-coplanar state also has the diagonal (density) and off-diagonal (BEC) orders simultaneously, it should be distinguished from the rigorous SS by the fact that the bosons on one of the three sublattices have no BEC order parameter. In other words, this state is partially disordered in the off-diagonal sector. Thus, the condensate flows on two sublattices avoiding the third, thus defining a honeycomb superlattice. We then refer to the $\pi$-coplanar state in the bosonic language as {\it superlattice} {\it superfluid}. Thus the 0-$\pi$ transition of coplanar spin states is expected to be observed as a transition between the SS and superlattice-superfluid states in the optical-lattice quantum simulator. Since these two interesting phases exist for large easy-plane anisotropy, the required strength of the dipole-dipole interaction ($=J_z$) is relatively small compared to the hopping amplitude ($= |J|/2$), which is more advantageous than the conditions needed for the observation of the SS in the negative $J/J_z$ side~\cite{wessel-05,bonnes-11,zhang-11,OpticalLattice,yamamoto-12}.

\emph{Conclusions.}---
We have studied the quantum phases of the spin-1/2 triangular-lattice {\it XXZ} model under magnetic fields motivated by the latest experimental developments in magnetism and optical-lattice systems. Using the dilute Bose-gas expansion and the CMF+S method, we established the entire quantum phase diagram including the frustrated regime and found that a nonclassical ($\pi$-)coplanar state emerges for strong fields. This is due to a particular lifting mechanism of the classical continuous degeneracy into two first-order transitions. We suggest that the quantum phase transition to the $\pi$-coplanar state can be observed in the magnetization process of triangular-lattice antiferromagnets with large easy-plane anisotropy or in the corresponding optical-lattice system.

The authors thank Tsutomu Momoi, Tetsuro Nikuni, Nikolay Prokof'ev, Hidekazu Tanaka, and Hiroshi Ueda for useful discussions. I.D.~is supported by KAKENHI from JSPS Grants No. 25800228 and No. 25220711.

\onecolumngrid

\newpage 

\subsection{\large Supplementary Material for ``Quantum Phase Diagram of the Triangular-Lattice {\it XXZ} Model in a Magnetic Field''}
\renewcommand{\thesection}{\Alph{section}}
\renewcommand{\thefigure}{S\arabic{figure}}
\renewcommand{\thetable}{S\Roman{table}}
\setcounter{figure}{0}
\newcommand*{\citenamefont}[1]{#1}
\newcommand*{\bibnamefont}[1]{#1}
\newcommand*{\bibfnamefont}[1]{#1}

\def\bs{{\bf S}}
\def\bk{{\bf k}}
\def\bp{{\bf p}}
\def\bq{{\bf q}}
\def\bQ{{\bf Q}}
\def\b0{{\bf 0}}
\def\br{{\bf r}}
\def\vpa{V^{\parallel}}
\def\vpe{V^{\perp}}
\def\dag{\dagger}
\def\cM{{\cal M}}
\def\bra{\langle}
\def\ket{\rangle}
\def\bbra{\langle\!\langle}
\def\kket{\rangle\!\rangle}
\def\vev#1{\langle{#1}\rangle}
\def\emin{\epsilon_{\rm min}}
\def\non{\nonumber\\}
\renewcommand{\theequation}{S\arabic{equation}}
\renewcommand{\thesection}{\Alph{section}}
\renewcommand{\thefigure}{S\arabic{figure}}
\renewcommand{\thetable}{S\Roman{table}}

\subsection{A. Magnon Bose-Einstein condensate (BEC)} 
\begin{figure}[b]
	\includegraphics[scale=0.65]{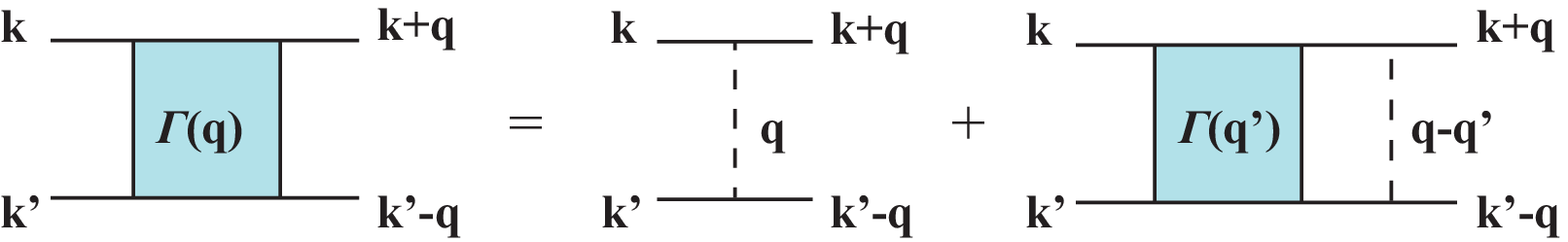}
\caption{{Ladder diagram included in}
Bethe-Salpeter equation (\ref{bseq}).
The filled squares and the dashed lines represent the full and the bare interaction respectively.}
\label{ladder}
\end{figure}

Using the hard-core boson map \cite{1956PThPh..16..569M} of spin 1/2 operators ($ S^-_i = a_i^\dag$, $ S^z_i=1/2 - a_i^\dag a_i $),  Eq.~(\ref{hamiltonian}) of the main text can be recast into an interacting-boson Hamiltonian, describing quantized spin waves (magnon), which in Fourier space reads, up to  constant terms,
\begin{align}
\hat{\mathcal{H}}= \sum_{\bk} [\epsilon(\bk) -\mu]\, \hat{a}^\dag_{\bk }\hat{a}_{\bk} + \frac{1}{2M} \sum_{\bk,\bk',\bq}  V(\bq ) \, \hat{a}^{\dag}_{\bk+\bq} \hat{a}^{\dag}_{\bk'-\bq} \hat{a}_{\bk' } \hat{a}_{\bk},
\label{boseham}
\end{align}
where $M$ is the number of lattice sites and
\begin{align}
\epsilon (\bk)  =& J  \left( \frac{3}{2}+ \nu(\bk) \right), \\
\nu(\bk) =& \cos ({k_x}) + \cos \left(\frac{k_x}{2}+\frac{\sqrt{3}
   k_y}{2}\right)+\cos
   \left(\frac{{k_x}}{2}-\frac{\sqrt{3}
   {k_y}}{2}\right), \\
V(\bq) =&  2 J_z \, \epsilon(\bq)  +U,\\
\mu =& 3 \left( J_z+\frac{J}{2} \right)- H \equiv H_s-H. \label{satf}
\end{align} 
Eq.~(\ref{satf}) contains the definition of the saturation field $H_s$. An on-site repulsive interaction $U>0$ has been introduced and will be eventually sent to infinity to implement the hard-core constraint (only zero or one boson per site are allowed). The single-magnon energy $\epsilon(\bk)$ admits two inequivalent minima ${\bf k}=\pm {\bf Q}\equiv\pm(4\pi/3,0)$ at the corners of the Brillouin zone. We thus expect $\hat{a}_\bQ$ and $\hat{a}_{-\bQ}$ to acquire a non-zero expectation value, denoted $\psi_{\pm {\bf Q}}\equiv \langle \hat{a}_{\pm{\bf Q}} \rangle$. We write the effective ground state energy, Eq.~(\ref{ene0}) in the main text, by expanding up to fourth order in $\psi_{\pm {\bf Q}}$ the  generating functional of one-particle-irreducible (1PI) correlation functions \cite{ZinnJustin:2002ru} and setting the frequency to zero:
\begin{align}
E_0/M\!\!=\!\!-\mu \left(|\psi_{{\bf Q}}|^2+|\psi_{-{\bf Q}}|^2\right) +\frac{{\mathit{\Gamma}}_1}{2}\left(|\psi_{{\bf Q}}|^4+|\psi_{-{\bf Q}}|^4\right)\!+\!{\mathit{\Gamma}}_2|\psi_{{\bf Q}}|^2|\psi_{-{\bf Q}}|^2
\label{eff}
\end{align}
The parameters $\mathit{\Gamma}_1$ and $\mathit{\Gamma}_2$ in \eqref{eff} are thus nothing but irreducible four-point functions calculated at appropriate external momenta. In the dilute limit $\mu \to 0$ ($H\to H_s$), formally they can be calculated in the ladder approximation \cite{beliaev1958application} solving the Bethe-Salpeter equation at zero total frequency and chemical potential
\begin{align}
 \mathit{\Gamma}(\bq;\bk,\bk') =  V(\bq)  - \frac{1}{M} \sum_{\bq' \in {\rm BZ} 
} 
 \frac{V(\bq-\bq')}{\epsilon(\bk+\bq')+\epsilon(\bk'-\bq')} \mathit{\Gamma}(\bq';\bk,\bk'),
\label{bseq}
\end{align}
which is diagrammatically depicted in Fig.~\ref{ladder}. Clearly $\mathit{\Gamma}_1= \mathit{\Gamma}({\bf 0};\bQ,\bQ)$ and $\mathit{\Gamma}_2= \mathit{\Gamma}({\bf 0};\bQ,-\bQ)+ \mathit{\Gamma}(-2\bQ;\bQ,-\bQ)$.

However, it is well-known that in two dimensions Eq.~\eqref{bseq} gives a logarithmically vanishing result at small  $\mu$ \cite{schick1971two,fisher1988dilute}, due to the singularities in the kernel. Moreover $\mathit{\Gamma}_1=\mathit{\Gamma}_2$ at leading order in $|\log\mu|^{-1}$ in this limit. To overcome this issue  we add a term to the Hamiltonian Eq.\eqref{hamiltonian}, which represents a small inter-layer coupling (with generic $XXZ$ anisotropy), namely 
\begin{align}
\delta\hat{\mathcal{H}}=J^\perp \!\sum_{\langle i,j\rangle_\perp} \!\left(\hat{S}_i^x\hat{S}_j^x+\hat{S}_i^y\hat{S}_j^y\right)\!+\!J^\perp_z\!\sum_{\langle i,j\rangle_\perp} \hat{S}_i^z\hat{S}_j^z.
\label{extraterm}
\end{align}

Even though this has a physical interpretation as a system of stacked weakly coupled triangular lattices (with the sum in Eq.\eqref{extraterm}  taken over nearest neighbors in the direction perpendicular to the layers), we essentially employ this extension to three dimensions as a regularization of Eq.~\eqref{bseq}. Upon calculating $\Gamma_1,\Gamma_2$ for progressively smaller values of $J^\perp/J$, we can extrapolate the result to the purely two-dimensional limit as explained below. A more extended discussion on this procedure, including more complicated models, can be found in \cite{Smarmorini2013}.
In the bosonic description we operate  the replacements $\epsilon(\bk)\to \epsilon(\bk) + J^\perp \cos k_z+|J^\perp|$, $V(\bq)\to V(\bq) +2J^\perp_z  \cos k_z$ and $H_s\to H_s+J^\perp_z +|J^\perp|$  and the two single-magnon minima become ${\bf k}=\pm {\bf Q}\equiv\pm(4\pi/3,0,0)$ (resp. $\pm(4\pi/3,0,\pi))$ for $J^\perp <0$ (resp. $J^\perp>0$). We then basically follow the procedure of Ref.~\onlinecite{Snikuni-95}. First  we 
integrate Eq.~\eqref{bseq} and find
\begin{align}
 \bra \mathit{\Gamma}(\bq;\bk,\bk') \ket =
 2U \left(1 - \frac{1}{M} \sum_{\bq' \in {\rm BZ}} \frac{\mathit{\Gamma}(\bq';\bk,\bk')}{\epsilon(\bk+\bq')+\epsilon(\bk'-\bq')} \right),
\label{bsequ}
\end{align}
where $\langle\ldots \rangle=(1/M)\sum_{\bq\in BZ}(\ldots)$. This can be used to eliminate $U$ from Eq.~\eqref{bseq}.  It is then possible to take the $U\to \infty$ limit and obtain
\begin{align}
1 - \frac{1}{M} \sum_{\bq' \in {\rm BZ}} \frac{\mathit{\Gamma}(\bq';\bk,\bk')}{\epsilon(\bk+\bq')+\epsilon(\bk'-\bq')} =0.
\label{bsequinf}
\end{align}
Let us define the two even functions $\widetilde{\mathit{\Gamma}}_1(\bq)=\mathit{\Gamma}(\bq;\bQ,\bQ)$ and $\widetilde{\mathit{\Gamma}}_2({\bq})= \mathit{\Gamma}(-\bQ+\bq;\bQ,-\bQ)+ \mathit{\Gamma}(-\bQ-\bq;\bQ,-\bQ)$. In the end $\mathit{\Gamma}_1=\widetilde{\mathit{\Gamma}}_1({\bf 0})$ and $\mathit{\Gamma}_2=\widetilde{\mathit{\Gamma}}_2(\bQ)$. By taking the ansatz
\begin{align}
\widetilde{\mathit{\Gamma}}_\alpha(\bq)=  \bra \widetilde{\mathit{\Gamma}}_\alpha \ket +  J_z A_\alpha \nu(\bq)  +J^\perp_z B_\alpha \cos {q_z}, \quad \alpha=1,2,
\label{ansatz}
\end{align}
and defining
\begin{align}
& {\bf T}(\bq) = (1,\nu(\bq), \cos q_z )^T, \\
&\tau_{ij} (\bk,\bk') = \frac{1}{M} \sum_{\bq'\in BZ} \frac{T_i(\bq') T_j(\bq')}
{\epsilon\left(\frac{\bk+\bk'}{2}+\bq'\right) +\epsilon\left(\frac{\bk+\bk'}{2}-\bq'\right)}, \\
&\tau^1=\tau(\bQ,\bQ), \quad \tau^2=\tau(\bQ,-\bQ),
\label{tau}
\end{align}
Eq.~\eqref{bsequinf} and Eq.~\eqref{bseq} [after the substitution Eq.~\eqref{bsequ}] can be reduced to the linear algebraic system
\begin{align}
& \left(
\begin{array}{ccc}
\tau^\alpha_{11} & J_z  \tau^\alpha_{12} & J^\perp_z \tau^\alpha_{13} \\
\tau^\alpha_{21} &  1+ 2 J_z  \tau^\alpha_{22} & 2 J^\perp_z \tau^\alpha_{23} \\
\tau^\alpha_{31} & 2  J_z  \tau^\alpha_{32} & 1+  2 J^\perp_z\tau^\alpha_{33} 
\end{array}
\right)
\left(
\begin{array}{c}
\langle \widetilde{\mathit{\Gamma}}_\alpha \rangle \\
A_\alpha \\
B_\alpha\\
\end{array}
\right)
= {\bf N}^\alpha \\
&{\bf N}^1=(1,2,2)^T,\quad {\bf N}^2=(2,-2,4)^T.
\end{align}
\begin{figure}[bt]
	\includegraphics[scale=0.61]{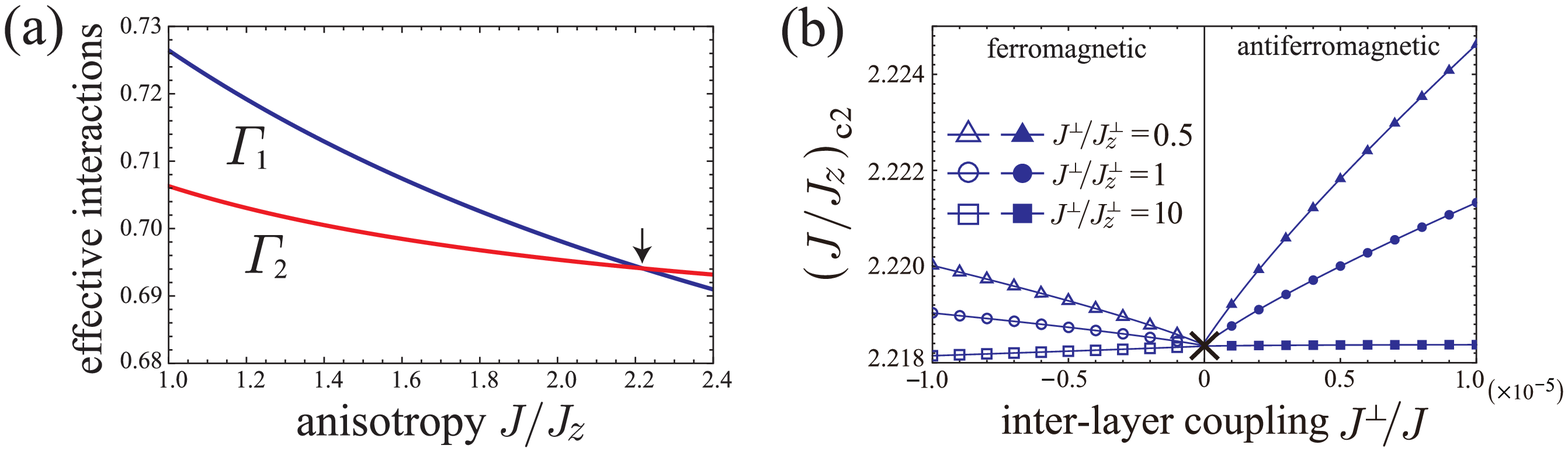}
\caption{(a) $\mathit{\Gamma}_{1,2}$ calculated at fixed $J^\perp/J=10^{-6}$ and $J^\perp/J^\perp_z=1$ as a function of the anisotropy $J/J_z$. (b) Extrapolation of the critical point $(J/J_z)_{c2}$; here we illustrate the independence on the type of inter-layer coupling by displaying both ferromagnetic and antiferromagnetic $J^\perp$ and $J^\perp/J^\perp_z=0.5,1,10$.}
\label{figs2}
\end{figure}
A sample calculation for small fixed $J^\perp/J$ is shown in Fig.~\ref{figs2}(a).
Eq.~\eqref{eff} is minimized by a single-mode BEC ($|\psi_{{\bf Q}}| =\mu/\mathit{\Gamma}_1$ and $|\psi_{-{\bf Q}}|= 0$ or vice versa)  for $\mathit{\Gamma}_1<\mathit{\Gamma}_2$ and by a two-mode BEC ($|\psi_{{\bf Q}}|=|\psi_{-{\bf Q}}| =\mu /(\mathit{\Gamma}_1+\mathit{\Gamma}_2)$) for $\mathit{\Gamma}_1>\mathit{\Gamma}_2$. By using the hard-core boson map it is easy to check that these  states correspond to umbrella and coplanar states respectively in the spin language, all of which are three-sublattice structures. The phase boundary between them is identified by the condition $\mathit{\Gamma}_1=\mathit{\Gamma}_2$, which gives the critical anisotropy  $(J/J_z)_{c2}$. We calculate $(J/J_z)_{c2}$ for progressively small $J^\perp/J$ (down to $J^\perp/J=10^{-6}$) and extrapolate for $J^\perp/J\to 0$ (see Fig.~\ref{figs2}(b), which includes the data displayed in Fig.~\ref{fig3} of the main text). Independently of the sign of $J^\perp$ (ferromagnetic or antiferromagnetic) and for any $J^\perp_z/J^\perp$ we reach the same limiting value $(J/J_z)_{c2}=2.218$, which is in excellent agreement (to within 0.1\%) with the CMF+S value $(J/J_z)_{c2}=2.220$.

\begin{figure}[h!]
\includegraphics[scale=0.7]{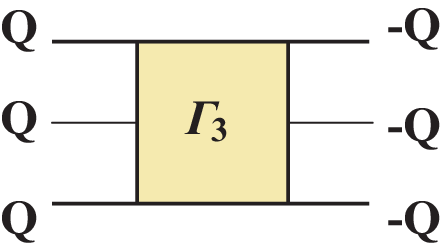}
\caption{The full three-magnon process that stabilizes either the 0- or the $\pi$-coplanar states.}
\label{threemag}
\end{figure}

The question of the other critical point $(J/J_z)_{c1}$ between the two different coplanar states can be formulated in the magnon theory as follows. As seen above, for $\mathit{\Gamma}_1>\mathit{\Gamma}_2$, $|\psi_{{\bf Q}}|=|\psi_{-{\bf Q}}| \equiv \psi \neq 0$ and the relative phase of the two condensates $\phi=\arg (\psi_{{\bf Q}}/\psi_{-{\bf Q}})$ is undetermined at the level of Eq.~\eqref{eff}. However, the three-particle process in Fig.~\ref{threemag} (remember that $3\bQ\simeq \mathbf{0}$ for the triangular lattice) will add a higher-order term $\delta E_0/M=2\mathit{\Gamma}_3 \psi^6 \cos 3\phi$ that can stabilize the relative phase. Namely $\phi=0, 2\pi/3$, or $4\pi/3$ for $\mathit{\Gamma}_3<0$ and  $\phi=\pi/3,\pi$, or $5\pi/3$ for $\mathit{\Gamma}_3>0$, from which we have chosen the terminology ``0-coplanar'' and ``$\pi$-coplanar''. Note that the three choices simply correspond to sublattice exchange.  The practical calculation of $\mathit{\Gamma}_3$ is, however, still an open problem, even in the dilute approximation; specifically at present it is not clear how  to sum up all the contributing diagrams.

\subsection{B. Comparison of the Transition Point at $H=0$ for $J/J_z<0$} 
The triangular-lattice spin-1/2 {\it XXZ} model in the unfrustrated regime ($J/J_z<0$) has been studied in the context of bosonic systems with the equivalent hardcore Bose-Hubbard model with nearest-neighbor (NN) interaction. The {\it XY} and Ising exchange interactions $J$ and $J_z$ are translated as twice the hopping amplitude $-2t$ and the NN interaction strength $V$, respectively. In this section, we give more quantitative evidence that our CMF+S method is consistent with the above studies. We focus on the transition between uniform and three-sublattice coplanar states at $H=0$ (see Fig.~1), which is the subject of most of the previous works. 
This transition is regarded as the superfluid-supersolid transtion in the bosonic language. We denote the phase transition point by $(J/J_z)_0$. In Table~\ref{table1}, we summarize the values of the transition point [for convenience we use $-(J/J_z)_0/2=(t/V)_0$] obtained by the single-site mean-field (MF) theory~\cite{Smurthy-97}, the quantum Monte-Carlo (QMC) simulations~\cite{Swessel-05,Sheidarian-05,Sbonnes-11}, a variational wave function approach with Monte Carlo optimization (VMC)~\cite{Sheidarian-10}, and our large-size cluster mean-field method combined with a scaling scheme (CMF+S). It can be seen that our CMF+S result is in good agreement with the other numerical data. 
\begin{table*}[hbt]
\caption{\label{table1}The transition point $(J/J_z)_0$ between the uniform and three-sublattice phases at $H=0$ from diffrent methods.}
\begin{ruledtabular}
\begin{tabular}{lccccccc}
~~ & MF~\cite{Smurthy-97} & QMC~\cite{Swessel-05} & QMC~\cite{Sheidarian-05} & VMC~\cite{Sheidarian-10} & QMC~\cite{Sbonnes-11} & CMF+S (present)~~ \\ \hline 
\rule[0pt]{0pt}{10pt}
~~ $-(J/J_z)_0/2$ & 0.25& 0.115 & $1/8.9\approx 0.112$ & $1/9.4\approx 0.106$ & 0.1108(2) [0.1105(3)]\footnote{The two values are obtained from the scaling of different quantities (see Ref.~\cite{Sbonnes-11}). } & 0.119~~ 
\end{tabular}
\end{ruledtabular}
\end{table*}

\subsection{C. The Square-Lattice {\it XXZ} Model}

\begin{figure}[tb]
\includegraphics[scale=0.2]{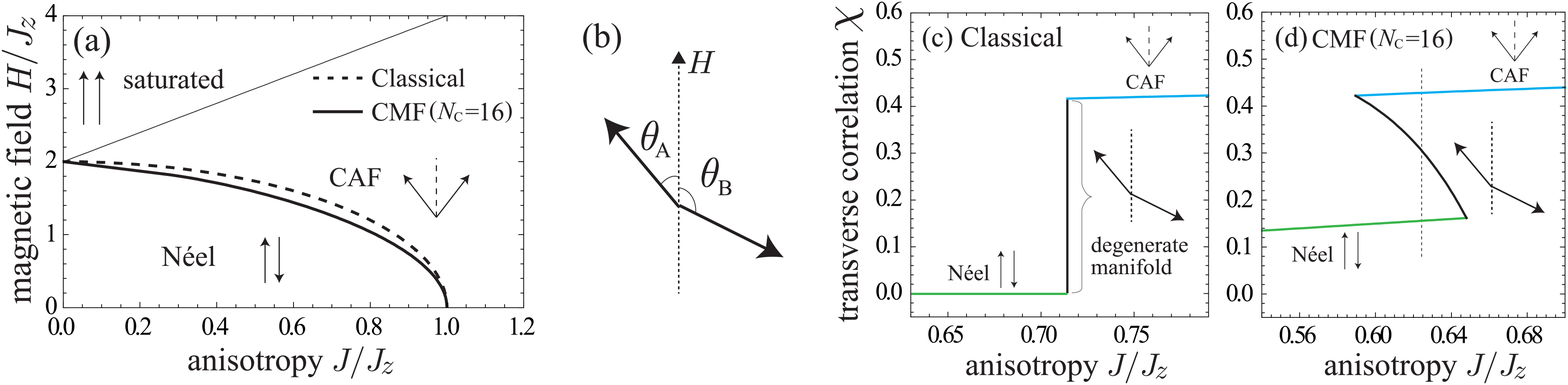}
\caption{\label{figSq} (color online).
(a) Ground-state phase diagram of the spin-1/2 square-lattice {\it XXZ} model obtained by the classical-spin (dashed curve) and CMF ($N_{\rm C}=16$; solid curve) analyses for $J_z>0$. The strength of the saturation field is not affected by the quantum effects. 
(b) General spin structure of a two-sublattice coplanar state. 
(c) The classical solution of $\chi$ as a function of $J/J_z$ for $H/J_z=1.4$. (d) The quantum degeneracy lifting obtained by the CMF analysis ($H/J_z=1.4$). The vertical dashed lines marks the first-order transition point determined by the Maxwell construction. 
}
\end{figure}

In the main text, we found that the lifting of the classical continuous degeneracy due to quantum fluctuations leads to the emergence of the new $\pi$-coplanar state. In order to elaborate that this degenercy-lifting mechanism is qualitatively different from the known examples, we briefly review the quantum effects on the ground states of the square-lattice {\it XXZ} model with the same Hamiltonian given by Eq. 1 of the main text. In Fig.~\ref{figSq}(a), we show the ground-state phase diagram obtained by the classical (MF) approximation~\cite{Sbruder-93,Spich-98} and the CMF ($N_{\rm C}=4\times4=16$) calculation. It has been pointed out in Ref~\cite{Sholtschneider-07} that there is an accidental coutinuous degeneracy of classical ground states along the spin-flop transition boundary between the N\'eel and canted antiferromagnetic (CAF) phases: $H_{\rm c}=2\sqrt{J_z^2-J^2}$. The degeneracy manifold consists of two-sublattice (checkerboard) coplanar states (called ``biconical'' states in Ref~\cite{Sholtschneider-07} or supersolid in the bosonic language~\cite{Smatsuda-70}) characterized by two classical spin angles $\theta_{\rm A}$ and $\theta_{\rm B}$ [Fig.~\ref{figSq}(b)] with the following constraint:
\begin{align}
\cos \theta_{\rm B}=\frac{\sqrt{J_z^2-J^2}-J_z\cos \theta_{\rm A}}{J_z-\sqrt{J_z^2-J^2}\cos \theta_{\rm A}}
\end{align}
It is well known that the classical degeneracy is lifted by the quantum flactuations and the spin-flop transition becomes a conventional first-order transition with a finite hysteresis region and the level clossing of the energy at the transition point~\cite{Sholtschneider-07,Skohno-97,Sbatrouni-00,Syunoki-02}. The previous QMC studies have clearly demonstrated that all the intermediate states are destabilized and only one first-order transition appears in the magnetication process of the spin-1/2 square-lattice {\it XXZ} model (compare, e.g., Fig.~3 and Fig.~4 of Ref.~\cite{Skohno-97}) and in the bosonic counterpart of the model~\cite{Sbatrouni-00}. Here, we supplementally show the quantum degeneracy lifting behavior in the plane of the anisotropy $J/J_z$ versus the transverse NN correlation $\chi$ in order to see more directly the difference from the peculiar lifting mechanism we found in the present study for the frustrated triangular lattice. As shown in Fig.~\ref{figSq}(c), the classical solution curve sharply change across the spin-flop transition point, at which the intermediate degenerate ground states smoothly connect the N\'eel ($\theta_{\rm A}=0$, $\theta_{\rm B}=\pi$ or vice versa) and CAF ($\theta_{\rm A}$=$\theta_{\rm B}$) phases. Figure~\ref{figSq}(d) illustrates the degeneracy lifting by the quantum flactuations. In contrast to the triangular lattice case shown in Fig. 4 of the main text, the intermediate section of the solution curve is not separated and its slope just becomes negative at each point, which means that only one first-order transition is induced by the quantum effects. In this case, no new phase emerge in the quantum phase diagram. Note that, to stabilize the intermediate biconical (or supersolid) states for the square lattice, it is required to include in the Hamiltonian a certain additional term that breaks the classical continuous degeneracy, such as single-ion anisotropy (for higher spins $S\geq 1$)~\cite{Sholtschneider-07-2} and longer-range interactions~\cite{Smatsuda-70,Sliu-73,Scapogrosso-10,Syamamoto-12}. 
\begin{figure}[t]
\includegraphics[scale=0.3]{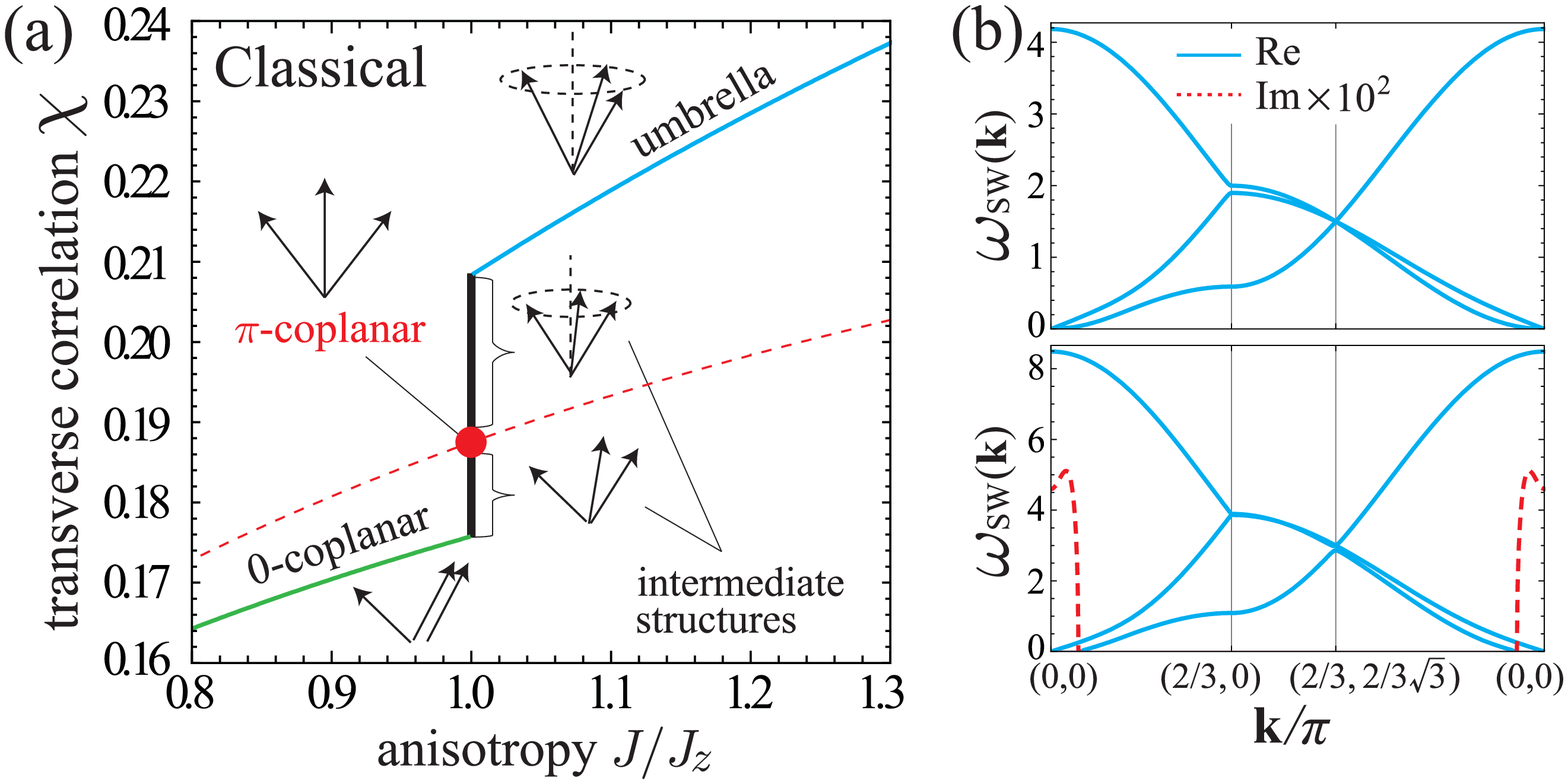}
\caption{\label{figTri} (color online). (a) The classical solution of the transverse correlation $\chi$ as a function of $J/J_z$ for $H/J_z=3$. The red dashed curve shows the solution of the $\pi$-coplanar state, which is unstable for $J/J_z\neq 1$. (b) Spin-wave excitations $\omega_{\rm SW}({\bf k})$ from the classical $\pi$-coplanar state for $J/J_z=1$ (upper) and $J/J_z=2$ (lower) at $H/H_{\rm s}=0.93$. }
\end{figure}
As we explained in the main text, the triangular-lattice model exhibits a quantitatively different quantum lifting, in which a specific ($\pi$-coplanar) state is chosen by the quantum flactuations and is stabilized in a finite region of the quantum phase diagram. We present here a discussion in terms of the spin-wave analysis as additional information. As stated in the main text, the $\pi$-coplanar state is just one of the infinite number of the degenerate classical ground states at $J/J_z=1$. However, as indicated by the red-dashed curve in Fig.~\ref{figTri}(a), the $\pi$-coplanar solution exists for any $J/J_z>0$ and $|H|<H_{\rm s}$ as a stationary point of the classical energy, unlike all the other intermediate states that interpolate between the 0-coplanar and umbrella states. The spin-wave excitation from the $\pi$-coplanar solution has an unstable imaginary mode for $J/J_z\neq 1$ as shown in Fig.~\ref{figTri}(b), which means that the classical $\pi$-coplanar state is stable only at $J/J_z=1$. The stabilization of the $\pi$-coplanar state in the quantum phase diagram can be interpreted as the result of the fact that the quantum fluctuations around the classical $\pi$-coplanar state significantly lower its energy for large easy-plane anisotropy.

\newpage 

\subsection{\large Erratum: Quantum Phase Diagram of the Triangular-Lattice {\it XXZ} Model in a Magnetic Field [Phys. Rev. Lett. {\bf 112}, 127203 (2014)]}
\renewcommand{\thesection}{\Alph{section}}
\setcounter{figure}{0}
\renewcommand{\thefigure}{E\arabic{figure}}
\renewcommand{\thetable}{E\Roman{table}}

\begin{figure}[b]
\includegraphics[scale=0.25]{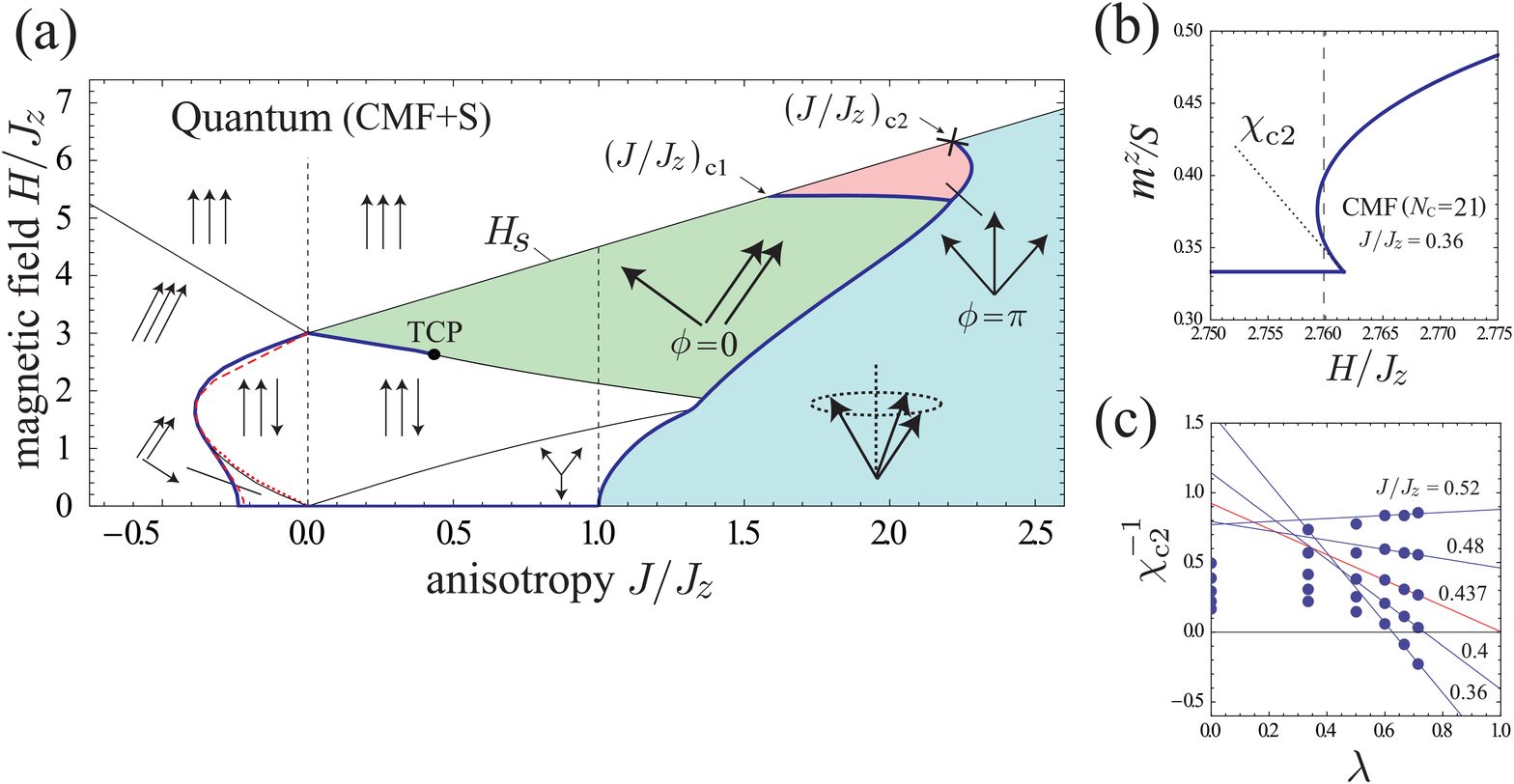}
\caption{\label{fig1E}
(a) Ground-state phase diagram of the spin-1/2 triangular-lattice {\it XXZ} model obtained by the CMF+S analyses ($J_z>0$). 
The thick blue (thin black) solid curves correspond to first- (second-) order transitions. The dot marks the tricritical point. The latest QMC data~\cite{Ewessel-05,Ebonnes-11} are shown by the red dashed (first-order) and dotted (second-order) curves. The symbol ($\times$) is the value from the dilute Bose gas expansion. (b) An example of the magnetization curve that exhibits a first-order transition. We show the magnetization $m^z$ divided by the saturation value $S=1/2$ as a function of the magnetic field $H/J_z$. The vertical dashed line marks the first-order transition point. (c) Cluster-size scalings of the CMF data for the inverse susceptibility $\chi_{{c}2}^{-1}$ just above the plateau. The lines indicate linear fits to the data obtained from the three largest clusters for each $J/J_z$. 
}
\end{figure}

We have realized that the transition between the up-up-down and $0$-coplanar states near the Ising limit  is actually of first order, although the hysteresis region is very narrow. In Fig.~\ref{fig1E}(a) we present the corrected quantum phase diagram, in which the boundary of the corresponding transition for $0<J/J_z< 0.437$ is replaced by a thick blue line. This minor correction does not affect the rest of the phase diagram and the main conclusions of the Letter~\cite{Eyamamoto-14}, including the novel degeneracy-lifting mechanism that gives rise to the new $\pi$-coplanar state.

Using the density matrix renormalization group (DMRG) method, a recent theoretical work has suggested that the transition between the up-up-down and $0$-coplanar states is of first order when $0<J/J_z\lesssim 0.4$~\cite{Esellmann-14}. Therefore, we reexamined the magnetization curve $m^z(H)=\sum_i \langle \hat{S}_i^z \rangle /M$ for small positive values of the anisotropy $J/J_z$. As shown in Fig.~\ref{fig1E}(b), the magnetization curve is three-valued in a very small but finite range of $H/J_z$ near the end point of the plateau ($H=H_{{c}2}$), which implies that the transition is of first order. When the anisotropy $J/J_z$ increases, the sign of the susceptibility $\chi_{{\rm c}2}\equiv dm^z/dH|_{H=H_{{ c}2}}$ just above the plateau changes from negative to positive, i.e., there is a tricritical point (TCP) where  the transition nature changes from first order to second order. In Fig.~\ref{fig1E}(c), we show the extrapolation of the inverse of $\chi_{{ c}2}$ with respect to the scaling parameter $\lambda$ for different values of $J/J_z$. The location of the TCP is estimated to be $(J/J_z)_{\rm TCP}\approx 0.437$, for which the extrapolated value of $\chi_{{ c}2}^{-1}$ is 0. This result is consistent with the recent report based on the DMRG calculation~\cite{Esellmann-14}. 


\begin{thebibliography}{99}
\bibitem{footnote}
The first-order nature of the transition between the uniform (superfluid) and three-sublattice (supersolid) states for $H\neq 0$ is overlooked in Ref.~\cite{wessel-05}. See Refs.~\cite{yamamoto-12,bonnes-11,zhang-11}.
\end{thebibliography}
\end{document}